\documentclass[prd,aps,amsmath]{revtex4}
\usepackage{graphicx}% Include figure files
\usepackage{amssymb,amsmath}
\newcommand{\be}{\begin{equation}}
\newcommand{\ee}{\end{equation}}
\newcommand{\bea}{\begin{eqnarray}}
\newcommand{\eea}{\end{eqnarray}}

\def\p1{\pi_1}

\begin{document}

\title{Apparent universality of semiclassical gravity in the far field limit}
\author{Paul R. Anderson$^{\rm a, \ b}$}
\altaffiliation{\tt anderson@wfu.edu}
\author{A. Fabbri$^{\rm b}$}
\altaffiliation{\tt afabbri@ific.uv.es}
\affiliation{${}^{a)}$Department of Physics, Wake Forest University,
Winston-Salem, NC 27109, USA,}
\affiliation{$^{b)}$ Departamento de F\'{\i}sica Te\'orica and IFIC,
Universidad de Valencia-CSIC, C. Dr. Moliner 50, Burjassot-46100, Valencia, Spain.}

\begin{abstract}

The universality of semiclassical gravity is investigated by
considering the behavior of the quantities $\langle \phi^2 \rangle$
and $\langle {T^a}_b \rangle$, along with quantum corrections to the
effective Newtonian potential in the far field limits of static
spherically symmetric objects ranging from stars in the weak field
Newtonian limit to black holes. For scalar fields it is shown that
when differences occur they all result from the behavior of a single
mode with zero frequency and angular momentum and are thus due to a
combination of infrared and s-wave effects. An intriguing
combination of similarities and differences between the extreme
cases of a Schwarzschild black hole and a star in the weak field
Newtonian limit is explained.

\end{abstract}

\pacs{04.62.+v, 04.70.Dy}

\maketitle

There is a long history of calculations of leading order quantum
corrections to the effective Newtonian potential using Feynman
diagrams for scalar and other loops in the graviton propagator
\cite{duff,donoghue,hamliu,aksh}. Another approach is to compute the
corrections to the Newtonian potential by solving the semiclassical
Einstein equations in the far field limit of a static spherically
symmetric object when the quantized fields are in the Boulware
state~\cite{boulware}. For a star in the weak field Newtonian limit
this has been done for both gravitons~\cite{mazz1} and massless
scalar fields with arbitrary coupling $\xi$ to the scalar
curvature~\cite{mazz}.  It has also been done for massless scalar
fields outside of a Schwarzschild black hole \cite{abf}.  In each
case it was found that the leading order correction depends only on
the mass of the source. The lack of dependence on the structure of
the source would seem to provide evidence of some type of
universality for semiclassical gravity in the far field limit.
However, examination of the results for massless scalar fields in
\cite{mazz,abf} shows that the corrections to the Newtonian
potential only agree in the cases of minimal ($\xi = 0$) and
conformal ($\xi=1/6$) coupling. Interestingly these are the only
cases we know of where the corrections for a massless scalar field
have been computed using the Feynman diagram technique
\cite{duff,hamliu}.

In Refs.\ \cite{abf,mazz} the quantities $\langle \phi^2 \rangle$
and $\langle {T^a}_b \rangle$ were also computed in the far field
limit.
 The results for a Schwarzschild black hole are \cite{abf,foot1}
\begin{subequations}
\bea & &\langle \phi^2 \rangle_{\rm Sch}  =  \frac{M}{24 \pi^2 r^3} \\
    & & \langle {T^a}_b \rangle_{Sch}  =  \frac{M}{240 \pi^2 r^5} \, {\rm diag} \left[ 4, 2, -3, -3 \right]
     + \frac{M}{8 \pi^2 r^5} (\xi - 1/6) \, {\rm diag} \left[2,-2,3,3 \right] \\
   & & (\Phi_q)_{\rm Sch} = \frac{M}{90 \pi r^3} \left[2 - 15 (\xi - 1/6) \right]
\eea \label{eq:bh-results}
\end{subequations}
\hspace{-0.25cm} where the quantum correction to the effective
Newtonian potential is denoted by $\Phi_q$. The differences between
the results for a star in the weak field Newtonian limit \cite{mazz}
and a black hole are
\begin{subequations}
\bea  \Delta \langle \phi^2 \rangle &=&  -  \frac{M}{4 \pi^2 r^3} \xi \\
      \Delta \langle {T^a}_b \rangle &=&   - \frac{3M}{4 \pi^2 r^5} \xi (\xi - 1/6)\, {\rm diag} [2,-2,3,3] \\
      \Delta (\Phi_q) &=& \frac{M}{\pi r^3} \xi (\xi-1/6)  \;.
\eea \
\label{eq:star-results}
\end{subequations}
 \hspace{-0.25cm}As  can be see from Eq.\ (\ref{eq:star-results}) there is complete agreement
 between the results only for $\xi = 0$.  For $\xi=1/6$ there is the puzzling situation that there is
no agreement for $\langle \phi^2 \rangle$ yet there is agreement for $\langle T^a_{\ b}\rangle$ and $\Phi_q$.
For all other values of $\xi$ there is complete disagreement between the two results.

To investigate the origin of these agreements and disagreements we
will use the formalism of Refs. \cite{paul,ahs} to compute the
quantities $\Delta \langle \phi^2 \rangle $ and $\Delta \langle
{T^a}_b \rangle$. The quantum correction to the Newtonian potential
$\Phi_q$ can be obtained by solving the semiclassical backreaction
equations and thus is closely tied to the value of  $\langle {T^a}_b
\rangle$.

Because of Birkhoff's theorem it is well known that the geometry
outside of any static spherically symmetric distribution of matter
(which for simplicity we have been referring to here as a star) is
the Schwarzschild geometry.  The general line element for a static
spherically symmetric spacetime can be written in the form \be ds^2
= -f(r) dt^2 + \frac{dr^2}{k(r)} + r^2 (d \theta^2 + \sin^2 \theta
\, d \phi^2) \;. \label{geme}\ee For Schwarzschild spacetime $ f = k
=1 - \frac{2 M}{r} $, with $M$ the mass of the black hole.  For a
static star, $f$ and $k$ will consist of different functions of $r$
inside the star but outside of it they will be the same as for a
black hole except that now $M$ is the mass of the star.

To compute the values of $\langle \phi^2 \rangle$ and $\langle
{T^a}_b \rangle$ outside of a static star we make use of the method
of Ref. \cite{paul, ahs} which is an extension of that developed in
Refs. \cite{caho}. We begin with $\langle \phi^2 \rangle$.  All
details are given in Ref. \cite{ahs}.

For a static spherically symmetric spacetime the Euclidean metric is
obtained using the relation $\tau = it$ with $\tau$ the Euclidean
time.  The Euclidean Green's function obeys the equation \be \Box_x
\, G_E(x,x') - \xi R(x) \, G_E(x,x') = -
\frac{\delta^4(x,x')}{\sqrt{g}}\ee with $\xi$ the coupling to the
scalar curvature $R$.  Formally the real part of the Euclidean
Green's function is equal to $\langle \phi^2 \rangle$ in the limit
the points come together.  Of course $G_E$ is divergent in this
limit. If renormalization is done using the method of point
splitting~\cite{dewitt,christensen} then one uses the
DeWitt-Schwinger expansion to compute the renormalization
counterterms which are then subtracted from $G_E$ before letting the
points come together. Writing $\langle \phi^2 \rangle_{unren} = G_E$
one has
 \be \langle \phi^2 \rangle_{\rm
ren} = \lim_{x' \rightarrow x} \left[ {\rm Re} \, \langle \phi^2
\rangle_{\rm unren}  - \langle \phi^2 \rangle_{\rm DS} \right] \;.
\label{eq:phi2ren}  \ee

For an arbitrary separation of the points the Euclidean Green's
function can be written as \be G_{\rm E}(x,x') = \frac{1}{4 \pi^2}
\int_0^\infty \, d \omega \cos[\omega(\tau-\tau')]
\sum_{\ell=0}^\infty (2 \ell+1) P_{\ell}(cos \gamma) C_{\omega \ell}
p_{\omega \ell}(r_<) q_{\omega \ell}(r_>)  \;.  \label{eq:GE} \ee
Here $P_\ell$ is a Legendre polynomial, $\cos \gamma = \cos \theta
\, \cos \theta' \, + \, \sin \theta \, \sin \theta' \, \cos(\phi -
\phi')$, and $C_{\omega \ell}$ is a normalization constant which is
determined by the Wronskian condition
 \be C_{\omega \ell} \left[p_{\omega \ell} \frac{d q_{\omega \ell}}{dr} - \frac{d p_{\omega \ell}}{dr} q_{\omega \ell}
\right] = - \frac{1}{r(r-2M)} \;. \label{eq:Wronskian} \ee The
radial modes $p_{\omega \ell}$ and  $q_{\omega \ell}$ obey the mode
equation \bea k \frac{d^2 S}{d r^2 } &+& \left[\frac{2 k}{r} +
\frac{k}{2 f} \frac{df}{dr} + \frac{1}{2} \frac{dk}{dr} \right]
\frac{dS}{dr}
   - \left[\frac{\omega^2}{f} + \frac{\ell(\ell+1)}{r^2} + \xi R \right]S = 0\  \label{eq:mode}
\eea where $S$ refers to one of the modes.  Note that $p_{\omega
\ell}$ is fixed by the boundary condition that it be regular either
at the event horizon of the black hole or the center of the star
while $q_{\omega \ell}$ is fixed by the boundary condition that it
be regular in the limit $r \rightarrow \infty$. Since the mode
equation has the same form outside of the star as it does outside
the event horizon of the black hole, $q_{\omega \ell}$ is the same
for both in these regions.

When the points are split in the time direction so that the radial
and angular points are the same then a high frequency version of the
WKB approximation for the radial modes can be substituted into Eq.\
(\ref{eq:GE}) to obtain the quantity $\langle \phi^2 \rangle_{\rm
WKBdiv}$.  This quantity has the same ultraviolet divergences as
$G_E$ in the limit that the points come together.  If it is both
added and subtracted from Eq.\ (\ref{eq:phi2ren}) then the result is
\be \langle \phi^2 \rangle_{\rm ren} = \langle \phi^2 \rangle_{\rm
numeric} + \langle \phi^2 \rangle_{\rm analytic} \;, \ee with \bea
\langle \phi^2 \rangle_{\rm numeric} &=& \langle \phi^2 \rangle_{\rm
unren} -
\langle \phi^2 \rangle_{\rm WKBdiv} \nonumber \\
\langle \phi^2 \rangle_{\rm analytic} &=& \langle \phi^2
\rangle_{\rm WKBdiv} - \langle \phi^2 \rangle_{\rm DS}  \nonumber
\eea  The quantity $\langle \phi^2 \rangle_{\rm analytic}$ has been
computed analytically~\cite{ahs} in an arbitrary static spherically
symmetric spacetime while $\langle \phi^2 \rangle_{\rm numeric}$
must usually be computed numerically.

While not truly local, $\langle \phi^2 \rangle_{\rm analytic}$ is
effectively local in the coordinate system we are using since it
just depends on the functions $f$ and $k$ and some of their
derivatives evaluated at a given value of the radial coordinate $r$.
Thus this piece is exactly the same outside the event horizon of a
black hole or outside the surface of a static spherically symmetric
star. The quantity $\langle \phi^2 \rangle_{\rm numeric}$ is
explicitly nonlocal and depends on the two radial mode functions
$p_{\omega \ell}$ and $q_{\omega \ell}$. In the region outside the
surface of a star it is \bea \langle \phi^2 \rangle_{\rm numeric}
&=& \frac{1}{4 \pi^2} \int_0^\infty d \omega \left[\sum_{\ell =
0}^\infty
 \left( (2 \ell+1) C_{\omega \ell} p_{\omega \ell}(r) q_{\omega \ell}(r) - \frac{1}{\sqrt{r(r-2M)}} \right)
 + \frac{\omega}{1-2M/r} \right]\ .
\label{eq:phi2num} \eea

Our goal is to derive Eq.\ (\ref{eq:star-results}) and to use the
results to understand the reasons for the agreements and
disagreements discussed above between the results for a black hole
and those for a static spherically symmetric star.  Thus we want to
compute the difference
\be \Delta \langle \phi^2 \rangle = \langle
\phi^2 \rangle_{\rm star} - \langle \phi^2 \rangle_{\rm black hole}
\label{eq:delphibhs}
\ee
in the region exterior to the surface of
the star in the case that the masses of the star and black hole are
equal. Since the spacetime geometry is the same in this region for
both the star and black hole, the renormalization counterterms must
also be the same. Thus $\Delta \langle \phi^2 \rangle$ is explicitly
finite and we expect that its value should be independent of the
renormalization scheme used provided that the Wald axioms are
satisfied~\cite{wald}. The same applies to the change in the
stress-energy tensor $\Delta \langle T_{ab} \rangle$ discussed
below.

Outside of the star $p_*$ and $q_*$ must be linear combinations of
the Schwarzschild modes $p_{\rm Sch}$ and $q_{\rm Sch}$. As
discussed above $q_*=q_{\rm Sch}$. However, since the boundary
condition for $p_*$ is fixed at the center of the star \be p_* =
\alpha p_{\rm Sch} + \beta q_{\rm Sch} \;. \label{eq:matching} \ee

The normalization constants $C_*$ and $C_{\rm Sch}$ are determined from the Wronskian
condition (\ref{eq:Wronskian}) using the appropriate mode functions.  Using (\ref{eq:matching})
it is easy to show that
\be \alpha C_* =  C_{\rm Sch} \;. \label{eq:Cstar}\ee
 Then combining Eqs.\ (\ref{eq:phi2num}) - (\ref{eq:Cstar}) one
  finds that
\be
 \Delta \langle \phi^2 \rangle = \frac{1}{4 \pi^2} \int_0^\infty d \omega \sum_{\ell = 0}^\infty
  (2 \ell+1) C_{{\rm Sch}} \frac{\beta_{\omega \ell}}{\alpha_{\omega \ell}}  q_{\omega \ell}^2 \;.  \label{eq:deltaphi2}  \ee
This formula is exact.
The normalization constant $C_{\rm Sch}$ can be determined from the asymptotic form of
the Schwarzschild mode functions.  At large $r$ these mode functions approach the
corresponding modes in flat space which are
\begin{subequations}
\bea p_{\rm flat} &=& \omega^{-\ell} i_\ell (\omega r)  \label{eq:pflat}  \\
 q_{\rm flat} &=& \omega^{\ell+1} k_\ell(\omega r) \label{eq:qflat}
\eea
\label{eq:pqflat}
\end{subequations}
\hspace{-0.25cm} with $i_\ell$ and $k_\ell$ modified spherical
Bessel Functions.  Substituting into Eq.\ (\ref{eq:Wronskian}) gives
\be C_{\rm Sch} = \frac{2}{\pi}  \label{eq:CSch} \ee for all values
of $\omega$ and $\ell$.

As long as the coefficients in the mode equation (\ref{eq:mode}) are
finite at the surface of the star the mode functions $p$ and $q$ and
their first derivatives are continuous there. Matching at the
surface $r = A$ gives
\begin{subequations}
\bea \alpha &=& \frac{p_*q_{\rm Sch}'-p_*'q_{\rm Sch}}{q_{\rm Sch}'p_{\rm Sch}-p_{\rm
Sch}'q_{\rm Sch}}  \\
     \beta &=& \frac{p_{\rm Sch}p_*'-p_{\rm Sch}'p_*}{q_{\rm Sch}'p_{\rm Sch}-p_{\rm
Sch}'q_{\rm Sch}}
\eea   \label{eq:alpha-beta}
\end{subequations}
\hspace{-0.25cm} where primes denote radial derivatives, and all
modes are evaluated at $r=A$. Substituting Eq.\
(\ref{eq:alpha-beta}) into Eq.\ (\ref{eq:deltaphi2}) shows that in
general there will be corrections due to the presence of the star
and that they will depend on the structure of the star since $p_*$
does.

In the large $r$ limit the metric goes to its flat space value and,
since the boundary conditions on the modes $q_{\omega \ell}$ are
imposed at $r = \infty$, these modes approach their flat space
values. Thus it suffices to use the flat space values for these
modes in Eq.\ (\ref{eq:deltaphi2}).  They are given in Eq.\
(\ref{eq:qflat}) and are of the form \be  q_{\rm flat} =
\frac{\pi}{2} \omega^{\ell + 1} e^{-\omega r} \left[
\frac{c_1}{\omega r}
     + ... + \frac{c_\ell}{(\omega r)^{\ell+1}} \right] \label{eq:qflat2}  \ee
with $c_1=1$ and the rest of the constants $c_i$ depending on the value of $\ell$.

Since the integral and sum in Eq.\ (\ref{eq:deltaphi2}) are finite
the order may be interchanged.  Then for a given value of $\ell$ the
major contribution to the integral comes from small values of
$\omega$.  A power series in inverse powers of $r$ can be generated
by successively integrating by parts provided that $\beta_{\omega
\ell}/\alpha_{\omega \ell}$ is a regular function of $\omega$ near
$\omega = 0$.  It is then found that the leading order contribution
comes only from the $\ell = 0$ integral and is \be \Delta \langle
\phi^2 \rangle = \frac{1}{16 \pi r^3} \frac{ \beta_{0 0}}{\alpha_{0
0}} \;. \label{eq:leading-order-Phi2} \ee This is a remarkable
simplification and works for all static spherically symmetric stars
provided only that the mode functions and their first derivatives
are continuous across the surface of the star, that $\beta_{\omega
\ell}/\alpha_{\omega \ell}$ is at least a $C^1$ function of $\omega$
at $\omega = 0$,
 and that the geometry outside the star is
the Schwarzschild geometry \cite{foot2}.

Using the result (\ref{eq:leading-order-Phi2}) we can show that
there is no correction at leading order if $\xi = 0$. First note
that if $\omega = \ell = \xi = 0$ the solution to the mode equation
(\ref{eq:mode}) which is regular on the event horizon of a black
hole or at the center of a spherically symmetric star is $p_{0 0} =
{\rm constant}$ and without loss of generality the constant can be
chosen to be $1$.  It follows immediately from Eq.\
(\ref{eq:matching}) that \be \label{xi0} \alpha_{0 0} = 1 \ , \ \ \
     \beta_{0 0} = 0\ .  \ee
Thus $\Delta \langle \phi^2 \rangle = 0$  if $\xi=0$. As shown below, this also implies that
 $\Delta \langle {T^a}_b\rangle = 0$ in this case, so
the leading order behaviors of $\langle \phi^2 \rangle$ and $\langle {T^a}_b\rangle$ do not
depend on the type of matter present so long as it is static and spherically symmetric.

Turning to the calculation of $\Delta \langle {T^a}_b \rangle$,
first note that, as for $\langle \phi^2\rangle_{analytic}$, we can
write \be \langle T^a_{\ b} \rangle =  \langle T^a_{\ b}
\rangle_{numeric} + \langle T^a_{\ b} \rangle_{analytic} \ee with
$\langle T^a_{\ b}\rangle_{analytic}$ the same outside the surface
of the star as outside the event horizon of a black hole of the same
mass. The nonzero components of $\langle {T^a}_b \rangle_{\rm
numeric}$ are made up of linear combinations of five mode sums
called $S_1$ ... $S_5$ in Ref. \cite{ahs}.  Ignoring the subtraction
terms, which are necessary for renormalization, but unnecessary for
the analysis below, these sums are of the form \be S_i = \frac{1}{4
\pi^2} \int_0^\infty d \omega \sum_{\ell = 0}^\infty (2 \ell + 1)
s_i\ , \ee with \bea s_1&=& \omega^2 C_{\omega \ell}p_{\omega \ell}
q_{\omega \ell},\;\;\; s_2=C_{\omega \ell}\frac{d p_{\omega
\ell}}{dr} \frac{d q_{\omega \ell}}{dr}, \;\;\; s_3=\left(\ell +
\frac{1}{2} \right)^2 C_{\omega \ell}p_{\omega \ell} q_{\omega
\ell},\nonumber
\\ s_4 &=& C_{\omega \ell} \left[ \frac{d p_{\omega \ell}}{dr}
q_{\omega \ell}
           + p_{\omega \ell} \frac{d q_{\omega \ell}}{dr} \right], \;\;\;\;\;\;\; s_5=C_{\omega
       \ell} p_{\omega \ell} q_{\omega \ell}\;.   \label{eq:Sn}\eea
One can find the changes in these sums $\Delta S_i$ for a static
star by substituting Eq.\ (\ref{eq:matching}) into the above
expressions, using Eq. (\ref{eq:Cstar}), and taking out the part
that would be there in pure Schwarzschild spacetime. This is exactly
what was done in deriving Eq. (\ref{eq:deltaphi2}), which is
 equivalent to $\Delta S_5$.  Operationally one simply replaces
 $C_{\omega \ell} p_{\omega \ell}$ by $C_{\rm Sch}\frac{\beta_{\omega \ell}}{\alpha_{\omega \ell}}q_{\omega \ell} $ in each of the
above expressions.

Then examination of Eq.\ (4.5) of Ref. \cite{ahs} shows that in the far field
limit outside of the star
\begin{subequations}
\bea  \Delta \langle {T^t}_t \rangle &=& \left(2 \xi + \frac{1}{2} \right)  \Delta S_1 + \left(2 \xi - \frac{1}{2} \right) \left( \Delta S_2
          +  \frac{1}{r^2} \Delta S_3
     -  \frac{1}{4 r^2} \Delta S_5 \right) \\
      \Delta \langle {T^r}_r \rangle &=& - \frac{1}{2}\Delta S_1 + \frac{1}{2} \Delta S_2
      -\frac{1}{2 r^2} \Delta S_3 + \frac{2\xi}{r}  \Delta S_4 + \frac{1}{8 r^2} \Delta S_5  \\
    \Delta \langle {T^\theta}_\theta \rangle &=& \left(2 \xi - \frac{1}{2} \right) \left( \Delta S_1 +  \Delta S_2 \right)
          + \xi \left(\frac{2}{r^2} \Delta S_3
      - \frac{1}{r} \Delta S_4   -  \frac{1}{2 r^2} \Delta S_5 \right)\ .
\eea
\label{eq:CeltaT}
\end{subequations}
  \hspace{-0.25cm} From the definitions of the modes sums $\Delta S_n$ it is immediately clear that in general \be \Delta S_4 = \frac{d \Delta S_5}{dr}  \;.\label{eq:S45}\ee

To go further note that since both in the case of a black hole and a star $\langle
T^a_{\ b}\rangle\sim 1/r^5$ examination of Eqs. (\ref{eq:CeltaT}) shows that these mode sums
in the far field limit go like

\be \Delta S_1 = \frac{C_1}{r^{5}}\ , \
 \Delta S_2 = \frac{C_2}{r^{5}}\ ,  \
 \Delta S_3 = \frac{C_3}{r^{3}}\ ,  \
 \Delta S_4 = \frac{C_4}{r^{4}}\ ,  \
 \Delta S_5 = \frac{C_5}{r^{3}}.
 \label{eq:delS}
\ee Substituting into Eq.\ (\ref{eq:S45}) gives \be C_4 = -3 C_5 \;.
\label{eq:C45} \ee If the radial derivative of $\Delta S_4$ is
computed and Eqs.\ (\ref{eq:mode}) and (\ref{eq:C45}) are used, then
to leading order one finds \be C_2 = - C_1 - C_3 + \frac{13}{4} C_5
\label{eq:C2a} \;. \ee Following the same procedure for $\Delta S_2$
gives \be C_2 = 5 C_1 + 3 C_3 - \frac{3}{4} C_5  \label{eq:C2b} \;.
\ee Eqs.\ (\ref{eq:C2a}) and (\ref{eq:C2b}) can be combined to
obtain the relations \be C_2 = \frac{1}{2} C_1 + \frac{9}{4} C_5 \;,
\;\;\; \ C_3 = -\frac{3}{2} C_1 + C_5\ .
       \label{eq:C23}
\ee This appears to be as far as one can go with this type of
constraint. However, it is possible to go further for the specific
leading order form of the mode sums.  From Eq.\ (\ref{eq:Sn}) it is
clear that the only difference between $\Delta S_1$ and $\Delta S_5$
is that the integrand of the former has an extra factor of
$\omega^2$.  When the same procedure is followed for computing the
leading order contribution to $\Delta S_1$ that was followed for
computing the leading order contribution for $\Delta \langle \phi^2
\rangle = \Delta S_5 $, the only difference is a factor of $1/2r^2$
arising from the difference between $\int_0^\infty d \omega
exp(-2\omega r)$ and $\int_0^\infty d \omega \omega^2 exp(-2\omega
r)$. Thus one finds that $ C_1 = C_5/2$. Using Eq.\
(\ref{eq:leading-order-Phi2}) to obtain $C_5$ the final result is
 \be C_5 = 2 C_1 = \frac{2}{5} C_2 = 4 C_3 = -\frac{1}{3} C_4
 = \frac{\beta_{00}}{16\pi \alpha_{00}}\ . \label{eq:Cn-answer} \ee

If Eqs.\ (\ref{eq:delS}) are substituted into Eqs.\ (\ref{eq:CeltaT}) and the relations (\ref{eq:Cn-answer}) are used
then it is found that to leading order
\be \Delta \langle {T^a}_b \rangle = \frac{3\beta_{00}}{16 \pi \alpha_{00} r^5} \left( \xi - 1/6 \right) { \rm diag}
\left[2, -2, 3,3 \right] \;.
\label{eq:leading-order-Tmn}  \ee
This immediately shows that for $\xi = 1/6$ the leading order contributions to the
components of $\Delta \langle {T^a}_b \rangle$ are zero
 even if the leading order contributions to the mode sums $\Delta S_1 ... \Delta S_5$ are
nonzero in this case. Moreover, from Eq. (\ref{xi0}) it follows that
when $\xi=0$,
% both $\Delta \langle \phi^2\rangle$ and
$\Delta \langle T^a_{\ b}\rangle $ vanishes.

As discussed in the introduction and shown explicitly in Eq.\
(\ref{eq:star-results}), there is agreement between the leading
order behaviors of $\langle \phi^2 \rangle$ in the case $\xi = 0$
and of $\langle T^a_{\ b} \rangle$ in the cases $\xi = 0$, $1/6$.
These results provide a mathematical explanation for these
agreements.

A potential weak point in the above derivation of the corrections to
$\langle \phi^2\rangle$ and $\langle T^a_{\ b}\rangle$ is the
assumption that $\beta_{00}/\alpha_{00}$ is a $C^1$ function of
$\omega$ at $\omega = 0$. We do not have a proof of this, but we
have numerically computed the full corrections for the case of
constant density stars. A previous calculation was done in
Ref.~\cite{hiscock} ``using the approximation developed by Page,
Brown and Ottewill''~\cite{pbo}.  We have obtained the leading order
behaviors of the full numerical corrections to both $\langle \phi^2
\rangle$ and $\langle {T^a}_b\rangle$ in the far field limit
\cite{foot3}. We have also computed these corrections by numerically
computing the quantity $\beta_{00}/\alpha_{00}$ which is given in
Eq.\ (\ref{eq:alpha-beta}), and substituting the result into Eqs.\
(\ref{eq:leading-order-Phi2}) and (\ref{eq:leading-order-Tmn}).  In
every case considered there is agreement between these calculations
to within the expected numerical error.

Next we consider the weak field limit. For the metric (\ref{geme})
this means that we can write $ f = 1 + \delta f$,
      $k = 1 + \delta k $
with $|\delta f|$, $|\delta k|$ and their derivatives assumed to be
small. If we write $p_* = p_{\rm flat} + \delta p$, to first order
in small quantities the mode equation for the case $\omega = \ell =
0$ becomes \be \frac{d^2 \delta p}{d r^2} + \frac{2}{r} \frac{d
\delta p}{dr} = -(\delta k) \frac{d^2 p_{\rm flat}}{d r^2} -
\left(\frac{2 \delta k}{r}
   + \frac{1}{2} \frac{d \delta f}{d r} + \frac{1}{2}\frac{d\delta k}{dr} \right) \frac{d
   p_{\rm flat}}{dr} + \xi R p_{\rm flat}
\;.  \label{eq:modeperturbation}  \ee Note that the derivative terms
on the right hand side vanish because $p_{\rm flat} = 1$. Then the
solution to the equation which has the correct boundary condition at
$r = 0$ is
  \be \delta p = \xi \int_0^r \frac{d r_1}{r_1^2} \int_0^{r_1} d r_2 r_2^2 R(r_2) \;.  \label{eq:deltap} \ee

In the weak field limit the geometry must be nearly flat everywhere.
Thus at the surface of the star
 \be p_{\rm Sch} = p_{\rm flat} + \delta p_{\rm Sch}\;, \;\;\;
 q_{\rm Sch} = q_{\rm flat} + \delta q_{\rm Sch}\ .
\label{eq:pqsch} \ee For the $\omega = \ell = 0$ mode $p_{\rm Sch} =
1$ so $\delta p_{\rm Sch} = 0$ \cite{candelas}. Substituting Eqs.\
(\ref{eq:deltap}) and (\ref{eq:pqsch}) into Eq.\
(\ref{eq:alpha-beta}) and then substituting the result into Eq.\
(\ref{eq:leading-order-Phi2}) one finds to first order that \be
\Delta \langle \phi^2\rangle = - \frac{\xi} {8 \pi^2 r^3} \int_0^A d
r r^2 R(r) \;. \label{eq:delphi3}  \ee

Einstein's equations can be used to write
$ R = 8 \pi (\rho - p_r - 2 p_\theta)  \label{eq:rho-p} $
with $\rho$ the density, $p_r$ the radial component of the pressure and $p_\theta$
its angular component.
  Thus the leading order correction in the weak field limit depends
on the equation of state of the matter that makes up the star.  In
the Newtonian limit the pressure becomes negligible compared to the
density and one finds \be \Delta \langle \phi^2\rangle = - \frac{\xi
M} {4 \pi^2 r^3} . \label{eq:delphi4}  \ee This shows that the
correction now depends only upon the mass of the star. Finally,
using this result for $\Delta S_5$ in Eqs.\ (\ref{eq:delS}) and
(\ref{eq:Cn-answer}) one finds \be \Delta \langle {T^a}_b \rangle =
-\frac{3 M}{4 \pi^2 r^5} \xi \left(\xi - 1/6 \right) {\rm
diag}[2,-2,3, 3] \label{eq:Tmn-Newtonian} \ee Comparison with Eqs.\
(\ref{eq:star-results}) shows that in the weak field Newtonian limit
our results (\ref{eq:delphi4}) and (\ref{eq:Tmn-Newtonian}) are in
complete agreement with those of Ref.\ \cite{mazz}.

One can ask what would happen for other fields.  Massless spin $1/2$
and spin $1$ fields are conformally invariant so it is reasonable to
expect that the apparent conformal symmetry, which apparently causes
the leading order terms in the stress-energy to be the same for a
star or black hole for the conformally invariant scalar field, will
lead to the same result for these fields.  Since gravitons are not
conformally invariant it is not clear what the relationship between
the stress-energy of gravitons outside of a star and outside of the
event horizon of a black hole will be.

For massive fields the mass provides an infrared cutoff.  If we
consider massive scalar fields with arbitrary curvature couplings
the derivation up to Eq.\ (\ref{eq:deltaphi2}) is the same. However,
in Eqs.\ (\ref{eq:pqflat}) and (\ref{eq:qflat2}) $\omega$ is
replaced by $ \Omega = \sqrt{\omega^2 + m^2}$, with $m$ the mass of
the field.  It is easy to show that in this case the corrections are
exponentially suppressed.  One would expect the same behavior for
massive fields of higher spin.

Our results show that, in general, there is a distinction between the far field limit for a strongly gravitating source of matter,
a weakly gravitating source of matter, and matter which can be treated in the Newtonian limit, even if the leading order result for the latter case
depends only upon the mass of the star.  Nevertheless in most cases of physical interest the leading order contributions to
the stress-energy for scalar fields and probably for fields of higher spin (except possibly gravitons) are independent of the type of matter
or its configuration so long as it is spherically symmetric.
This leads to an apparent universality
in the quantum corrections these fields make to the
effective Newtonian potential.

We thank A. Akhundov, R. Balbinot, S. Fagnocchi, F. D. Mazzitelli,
J. Navarro-Salas, and M. Rinaldi for discussions. We would also like
to thank the Physics Department at the University of Bologna for
hospitality.
 This work was supported in part by the National Science Foundation
under grant number PHY05-56292, the Spanish grant
FIS2005-05736-C03-03 and the EU Network MRTN-CT-2004-005104. P.R.A.
acknowledges the Spanish Ministerio de Educaci\'on y Ciencia for
financial support.


\begin{references}
\bibitem{foot1} The computations of \cite{abf}
involve the quantity $\langle \phi^2\rangle$, although its value is not reported
in that paper.
\bibitem{foot2} It seems likely that this could
be extended to static spherically symmetric stars with a nonzero
electric charge, in which case the geometry outside the star would
be the Reissner-Nordstr\"om geometry.
\bibitem{foot3} As was mentioned in Ref. \cite{abf} the analytic approximations in \cite{pbo, ahs} give a
leading order behavior of the stress energy for the Boulware state
in Schwarzschild spacetime proportional to $M^2/r^6$.
\end{references}
\end{document}